\documentclass[%
 reprint,
 amsmath,amssymb,
 aps,
]{revtex4-2}
\usepackage{graphicx}
\usepackage{dcolumn}
\usepackage{bm}
\usepackage[version=3]{mhchem}
\usepackage{textcomp}
\usepackage{xcolor}
\definecolor{niceblue}{RGB}{55,126,184}
\definecolor{nicered}{RGB}{228,26,28}
\definecolor{drkgreen}{RGB}{0,120,90}
\usepackage[colorlinks,breaklinks,bookmarks=true,citecolor=niceblue,linkcolor=nicered,urlcolor=niceblue]{hyperref}

\usepackage{hyperref}

\begin{document}

\preprint{APS/123-QED}

\title{$J_{\rm{eff}}$ = 1/2 Hyperoctagon Lattice in Cobalt Oxalate Metal-Organic-Framework}

\author{Hajime Ishikawa}
\email{hishikawa@issp.u-tokyo.ac.jp}
\author{Shusaku Imajo}
\author{Hikaru Takeda}
\author{Masafumi Kakegawa}
\author{Minoru Yamashita}
\author{Jun-ichi Yamaura}
\author{Koichi Kindo}
\affiliation{%
Institute for Solid State Physics, University of Tokyo, Kashiwa, Chiba, 277-8582, Japan 
}%

\date{\today}

\begin{abstract}
	We report the magnetic properties of a cobalt oxalate metal-organic-framework featuring the hyperoctagon lattice. Our thermodynamic measurements reveal the $J_{\rm{eff}}$ = 1/2 state of the high-spin \ce{Co^2+} (3$\textit{d}^{7}$) ion and the two successive magnetic transitions at zero field with two-stage entropy release. $^{13}$C-NMR measurements reveal the absence of an internal magnetic field in the intermediate temperature phase. Multiple field-induced phases are observed before full saturation at around 40 T. We argue the unique cobalt oxalate network gives rise to the Kitaev interaction and/or a bond frustration effect, providing an unconventional platform for frustrated magnetism on the hyperoctagon lattice.
\end{abstract}

\maketitle

\paragraph*{Introduction}
Spin frustration is a key concept in condensed matter physics \cite{lacroix2011introduction}. The geometrical frustration in non-bipartite triangular and kagome lattices hinders N{\'e}el order of the constituent spins, resulting in an exotic ground state such as a quantum spin liquid (QSL) \cite{zhou2017quantum}. In the last decade, a honeycomb lattice with bond-dependent anisotropic interaction (Kitaev interaction) has become the paradigm of QSL study \cite{kitaev2006anyons}. Despite the bipartite structure, an exactly solvable QSL is realized where fractional excitations described by itinerant Majorana fermions and localized $Z_{\rm{2}}$ fluxes emerge \cite{trebst2022kitaev,motome2020hunting}. 

Kitaev interaction is proposed to appear between magnetic ions with spin-orbital entangled total angular momentum $J_{\rm{eff}}$ = 1/2 \cite{jackeli2009mott}. When the metal-ligand octahedra share an edge, two superexchange paths cause a quantum interference, which may cancel the isotropic Heisenberg interaction and leaves the Kitaev interaction. This Kitaev interaction is suggested to appear in 4\textit{d} ruthenates and 5\textit{d} iridates with $\textit{d}^{5}$ electron configuration, leading to extensive experimental investigations \cite{trebst2022kitaev,motome2020hunting}. Recently, 3\textit{d} cobaltate and 4\textit{f} systems with $J_{\rm{eff}}$ = 1/2 ions are also suggested as the Kitaev candidate materials. \cite{sano2018kitaev,liu2018pseudospin,motome2020materials}.

QSLs may emerge in the three-dimensional (3D) frustrated lattices such as pyrochlore \cite{canals2000quantum} and hyperkagome \cite{okamoto2007spin} lattices. It has been known that the Kitaev honeycomb model can be extended into various 3D tri-coordinated lattices \cite{trebst2022kitaev, eschmann2020thermodynamic}, providing a rich platform for QSL physics. However, due to the difficulty in realizing 3D variants of the honeycomb lattice, material explorations have been limited to 2D layered honeycomb structure \cite{trebst2022kitaev,xiao2019crystal,yao2020ferrimagnetism,zhong2020weak,songvilay2020kitaev,kim2021antiferromagnetic,wang2023single,halloran2023geometrical,daum2021collective,ishikawa2022sm}, except a few 3D iridates \cite{biffin2014noncoplanar,takayama2015hyperhoneycomb}.

\begin{figure}[tb]
	\includegraphics[width=8.6cm]{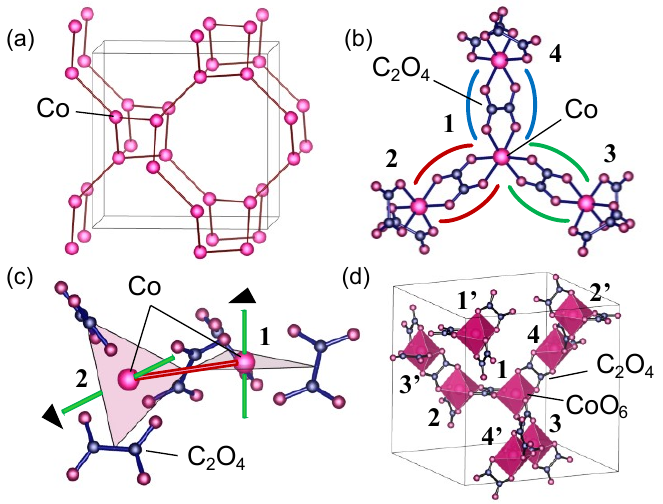}
	\caption{\label{fig:fig1} (a) Hyperoctagon lattice of Co in \ce{Co2(ox)3} network. All Co ions are crystallographically equivalent. (b) Local structure of \ce{Co2(ox)3} network seen from the trigonal axis of Co at the center. The numbers indicate the position in the unit cell as shown in (d). (c) Tilting of the planes perpendicular to the trigonal axis (pink triangles and green lines). The red line is shared by the neighboring planes. (d) \ce{Co2(ox)3} network in the unit cell. \ce{CoO6} octahedra labeled by different numbers have different trigonal axis directions. Figures are depicted by VESTA software \cite{momma2011vesta}}
\end{figure}

Metal-organic-framework (MOF), an emerging class of porous material, can potentially pave the way to study the magnetism in complex 3D lattices. We focus on the 3D lattice found in \ce{[(Me2NH2)3(SO4)]2[Co2(ox)3]} \cite{li2009d} (Fig.~\ref{fig:fig1}(a)): Me and ox indicate \ce{CH3} and \ce{C2O4}, respectively. The crystal structure features the interpenetrating \ce{[(Me2NH2)3(SO4)]2} and \ce{Co2(ox)3} networks. In the latter, \ce{Co^2+} (3$\textit{d}^{7}$) ions form the hyperoctagon lattice, which is cubic, chiral, and one of the 3D extensions of the honeycomb lattice \cite{hermanns2014quantum}. There are two superexchange paths between Co ions via an oxalate anion as in the case of the edge-shared octahedra (Fig.~\ref{fig:fig1}(b)). Indeed, realization of the Kitaev model in such an oxalate MOF with $J_{\rm{eff}}$ = 1/2 state of $\textit{d}^{5}$ ion is theoretically proposed. \cite{yamada2017designing}.

In this letter, we demonstrate the first experimental realization of $J_{\rm{eff}}$ = 1/2 hyperoctagon lattice in \ce{[(Me2NH2)3(SO4)]2[Co2(ox)3]}. We find multiple magnetic phases including an unusual intermediate temperature phase without magnetic order. Based on the structural considerations, we argue that these multiple magnetic phases are brought by the frustration arising from a Kitaev interaction in the nearest neighbor superexchange process via the oxalate anion and/or a bond frustration effect caused by the tilting of local trigonal axes in the \ce{Co2(ox)3} network (Fig.~\ref{fig:fig1}(c,d)). Our results demonstrate that the cobalt oxalate MOF serves as a platform for studying the frustrated magnetism on the hyperoctagon lattice.

\paragraph*{Results}
 Powder samples, including single crystals of $\sim 100$~$\mu$m size, were prepared by a solvothermal method \cite{li2009d}. The samples were handled in argon atmosphere to take care the slight hygroscopic nature. Phase purity was confirmed by the powder x-ray diffraction analysis using FullProf software \cite{rodriguez1993recent}: see supplemental material (SM) \cite{Supple}. Single crystal x-ray diffraction measurement was performed by a diffractometer with Mo-K$ \alpha$ radiation (Rapid II, Rigaku), revealing the crystal structure with the cubic space group $ \textit{I}{4_{1}32} $ as reported previously and a lattice constant $ \textit{a} $ = 15.454(1) \AA \space at room temperature. The Flack parameter was estimated to be zero, indicating the chiral crystal is obtained. 

\begin{figure}[tb]
	    \includegraphics[width=8.6cm]{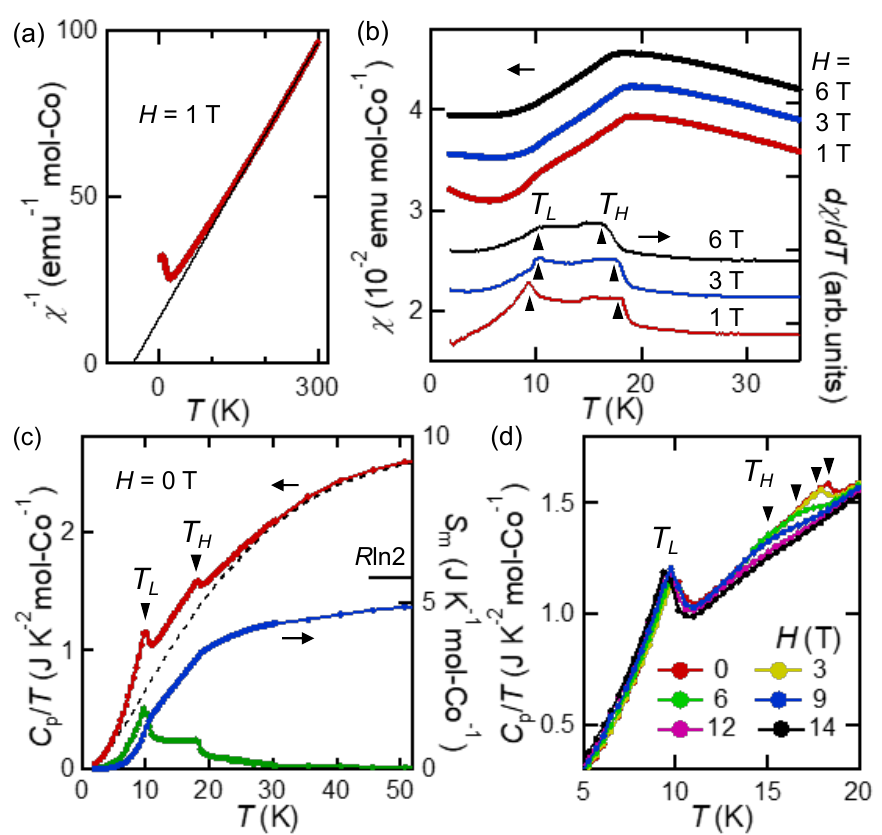}
    \caption{\label{fig:fig2} (a) Temperature dependence of inverse susceptibility of the powder sample at 1 T with a Curie-Weiss fit (solid line). (b) Temperature dependence of magnetic susceptibility (left axis) and the temperature derivative (right axis) at 1, 3, and  6 T. The data at 3 and 6 T is shifted for clarity. (c) Temperature dependence of specific heat divided by temperature (red, left axis) measured on the powder pellet at zero field. The black dashed line shows the estimated lattice contribution. Temperature dependence of magnetic specific heat (green, left axis) and magnetic entropy (blue, right axis) are also plotted. (d) Temperature dependence of specific heat divided by temperature at various magnetic fields.
    }
\end{figure}

Magnetization of the powder sample was measured by a SQUID magnetometer at 1.8--300 K (MPMS-XL, Quantum Design). The inverse magnetic susceptibility above 200 K at 1 T follows the Curie-Weiss law of $\chi(T) = C/(T \space+ \space\Theta)$ (Fig.~\ref{fig:fig2}(a)) with the Curie constant \textit{C} = 3.641(3) emu/mol-Co·K and the Weiss temperature $\Theta$ = 50.3(2) K, indicating the dominant antiferromagnetic interaction. From the Curie constant, the effective magnetic moment $\mu_{\rm{eff}}$ is estimated as $\mu_{\rm{eff}}$ = 5.40 $\mu_{\rm{B}}$, where $\mu_{\rm{B}}$ is the Bohr magneton. $\chi(T)$ at 1 T shows a peak at $T_{\rm{H}}$ = 18 K followed by a kink at $T_{\rm{L}}$ = 10 K (Fig.~\ref{fig:fig2}(b)). The anomalies at $T_{\rm{H}}$ and $T_{\rm{L}}$, which are more clearly visible in the temperature derivative $\textit{d$\chi$}$/$\textit{dT}$, suggest two successive antiferromagnetic transitions. We find that $T_{\rm{H}}$ is suppressed to lower temperatures by increasing the magnetic field, whereas $T_{\rm{L}}$ is slightly increased at 3 T and 6 T from that at 1 T.

Specific heat of the powder sample was measured by the relaxation method using a commercial apparatus (PPMS, Quantum Design). At zero magnetic field, the specific heat divided by temperature $C_p/T$ exhibits a kink at 18 K and a peak at 10 K (Fig.~\ref{fig:fig2}(c)) in accord with $T_{\rm{H}}$ and $T_{\rm{L}}$ in $\textit{$\chi$}$. We find that the peak at $T_{\rm{H}}$ is shifted to lower temperature by increasing the magnetic field (Fig.~\ref{fig:fig2}(d)), as observed in the field dependence of the anomaly of $\chi$ at $T_{\rm{H}}$. At 12 T, the peak at $T_{\rm{H}}$ is no longer visible by the broadening. In contrast, the peak at $T_{\rm{L}}$ is insensitive to the magnetic field up to 14 T. To estimate the lattice specific heat, we synthesized the nonmagnetic and isostructural Zn-analogue \cite{nagarkar2014two}. The lattice contribution in the Co-compound is estimated by multiplying a factor to the $C_p/T$ data of the Zn-compound so that the two data almost coincide at 50 K (Fig.~\ref{fig:fig2}(c)): the temperature scale of the Zn-compound is renormalized by considering the difference in the molecular weight. The magnetic specific heat is estimated by subtracting the lattice contribution and is integrated to obtain magnetic entropy $S_m$, which piles up to approximately 4.9 J/mol-Co$\cdot$K at 50 K. 
 
 \begin{figure}[tb]
 	\includegraphics[width=8.6cm]{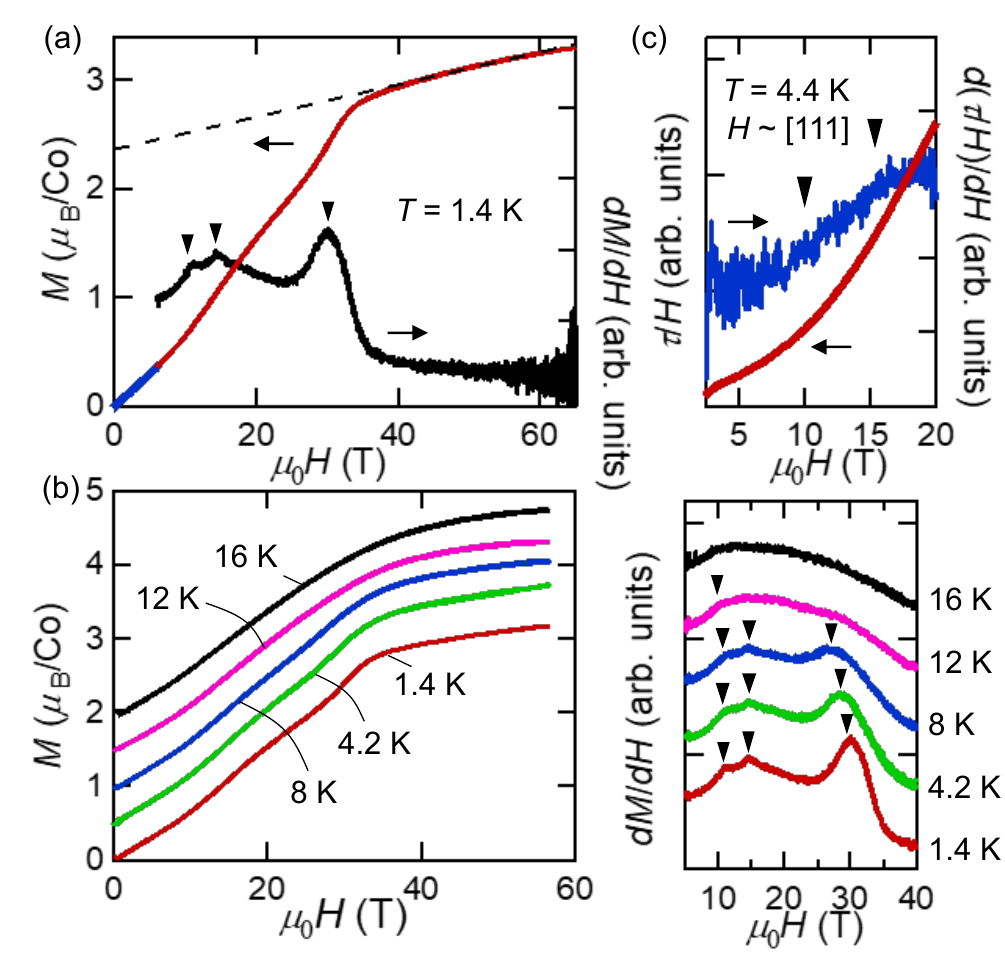}
 	\caption{\label{fig:fig3} (a) Magnetization curve measured on the powder sample in the pulsed magnetic field at 1.4 K (red) and by the SQUID at 1.8 K (blue). The field derivative of the pulsed field data is plotted in the right axis. The dashed line shows a linear fit above 40 T. (b) Magnetization curves at different temperatures (left) and their magnetic field derivatives (right): data is shown with offset. (c) Field dependence of the magnetic torque divided by the magnetic field (red) and its magnetic field derivative (blue) measured on one single crystal.}
 \end{figure}
 
 To investigate the magnetic phase diagram up to full saturation, we performed magnetization measurements of the powder sample by an induction method in the pulsed magnetic field up to 65 T. At 1.4 K, the magnetization curve exhibits anomalies below 20 T and a saturation at around 40 T (Fig.~\ref{fig:fig3}(a)). The magnetization gradually increases up to 65 T, which can be attributed to the Van Vleck paramagnetism. A linear fit to the magnetization curve above 40 T yields the slope of 0.015 $ \mu_{\rm{B}}$/T, which corresponds to 8.4 $\times$ $10^{-3}$ emu/mol-Co, and the saturation moment of 2.36 $ \mu_{\rm{B}}$/Co.
 
 In the magnetic field derivative of the magnetization at 1.4 K, clear anomalies are observed at around 11, 15, and 30 T. As raising the temperature, the anomalies are weakened, but visible up to around 12 K (Fig.~\ref{fig:fig3}(b)). To confirm that these anomalies at 11 T and 15 T are not caused by the inhomogeneous distribution of the crystal orientations in the powder sample, we measured the magnetization of one single crystal by detecting the magnetic torque divided by magnetic field ($\tau$/\textit{H}), which is proportional to \textit{M} by the magnetic anisotropy (Fig.~\ref{fig:fig3}(c)). The sample was attached on a microcantilever with the [111] axis perpendicular to the microcantilever plane, and the magnetic field was applied almost parallel to the [111] axis: see Fig.S1(a) in SM \cite{Supple}. As shown in Fig.~\ref{fig:fig3}(c), the field dependence of $\tau$/\textit{H} shows anomalies at 10 T and 15 T, which are observed as kinks in the field derivative. These anomalies in the torque measurement confirm the presence of successive anomalies in the magnetization.

 \begin{figure}[tb]
 	\includegraphics[width=8.6cm]{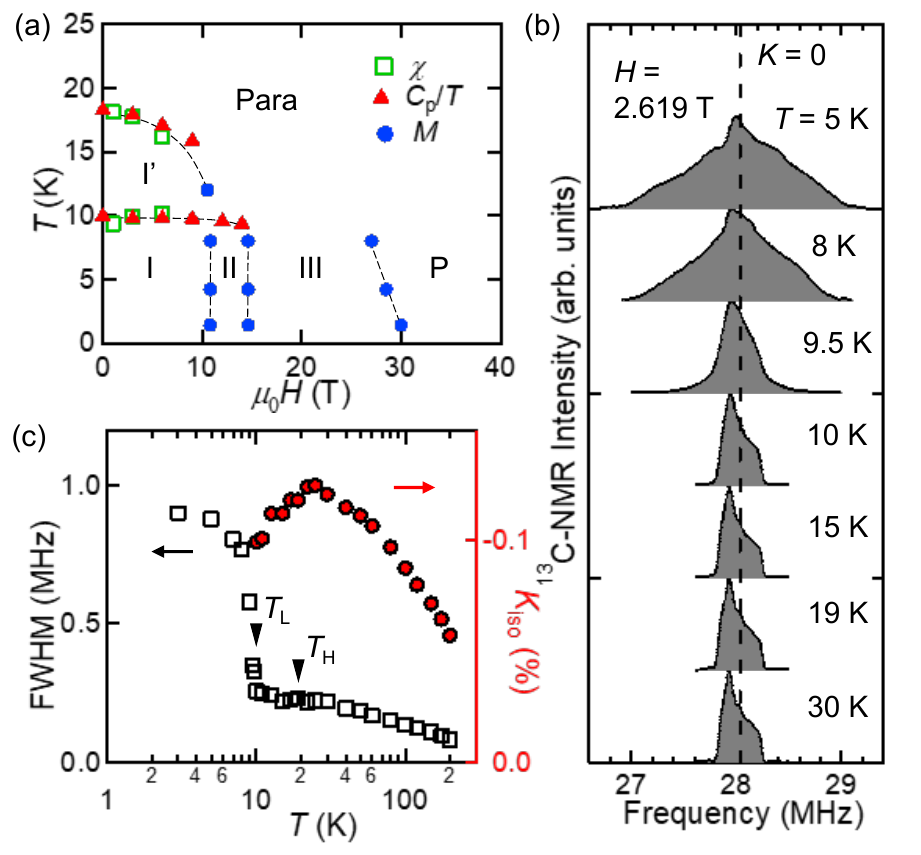}
 	\caption{\label{fig:fig4} (a) $H$--$T$ phase diagram determined by the anomalies observed in $\chi$ (squares), $C_p/T$ (triangles), and $\textit{M}$ (circles). Phases are labeled by I, I', II and III. Para and P indicate the paramagnetic and fully polarized states, respectively. (b) $^{13}$C-NMR spectra measured on the powder sample at 2.619 T at selected temperatures below 30 K. The dashed line indicates the frequency for the NMR shift $K = 0$. (c) Temperature dependence of the full width at half maximum of the spectra (black squares, left axis) and the isotropic part of the NMR shift (red circles, right axis).}
 \end{figure}
 
In order to identify the magnetic states in the phases I' and I, we conducted $^{13}$C ($I = 1/2$) nuclear magnetic resonance (NMR) measurements on the powder sample with $^{13}$C-enriched oxalate at 2.619 T. The positional relationship between the C in the oxalate ion and the neighboring Co ions makes the $^{13}$C-NMR a sensitive probe for detecting the appearance of an internal magnetic field \cite{Supple}. The NMR spectra (Fig.~\ref{fig:fig4}(b)) in the paramagnetic state above $T_{\rm{H}}$ exhibit the asymmetric powder pattern typical for the $I=1/2 \leftrightarrow -1/2$ transitions with anisotropic NMR shifts: see Fig.S2 in SM for details of the NMR shift analyses \cite{Supple}. Remarkably, there is no discernible difference in the NMR spectrum below $T_{\rm{H}}$, except for the decrease of the NMR shift corresponding to the decrease of $\chi$ (Fig.~\ref{fig:fig4}(c)). In contrast, the spectrum broadens below $T_{\rm{L}}$, indicating the appearance of internal magnetic fields formed by a magnetic order in the phase I. The temperature dependence of the spectrum width (Fig.~\ref{fig:fig4}(c)) clearly shows the temperature dependence of the internal magnetic field that starts to develop below $T_{\rm{L}}$. The absence of the internal field despite the sizable decrease of $\chi$ is an unusual feature of the phase I’.

\paragraph*{Discussion}
We first discuss the formation of the $J_{\rm{eff}}$ = 1/2 state. The effective magnetic moment $\mu_{\rm{eff}}$ = 5.40 $\mu_{\rm{B}}$ clearly exceeds the spin-only value of 3.87 $\mu_{\rm{B}}$ for \textit{S} = 3/2, indicating the large orbital contribution. Note that the $\mu_{\rm{eff}}$ falls within the range of the values reported in 2D honeycomb cobaltates (5--6 $\mu_{\rm{B}} $) \cite{yao2020ferrimagnetism, zhong2020weak, wang2023single}. The \textit{g}-value of 4.72 estimated from the saturation moment is close to 4.33 expected for the $J_{\rm{eff}}$ = 1/2 doublet of \ce{\textit{d}^7} ion in the ideal cubic octahedral crystal field \cite{abragam1970electron}. The slight enhancement of the \textit{g}-value may be attributed to the deviation from the ideal cubic crystal field that allows the mixing of the higher energy multiplets. Moreover, the magnetic entropy of 85\% of \textit{R}ln2 (Fig.~\ref{fig:fig2}(c)) observed below 50 K is significantly smaller than \textit{R}ln4 expected for \textit{S} = 3/2. From these results, we can safely conclude that $J_{\rm{eff}}$ = 1/2 doublet is formed in the \ce{Co^2+} ions.

The formation of the $J_{\rm{eff}}$ = 1/2 state is compatible with the trigonal distortion of the \ce{CoO6} octahedra in this compound. As shown in Fig.S1(b) in SM \cite{Supple}, the \ce{CoO6} octahedra in the \ce{Co2(ox)3} network are compressed along the trigonal axis, which is clearly the opposite situation that a large trigonal elongation of the octahedron results in the \textit{S} = 3/2 ground state by totally quenching the orbital angular momentum \cite{khomskii2014transition}. In fact, the formation of the $J_{\rm{eff}}$  = 1/2 doublet is observed in honeycomb cobaltates with similarly compressed \ce{CoO6} octahedra \cite{songvilay2020kitaev,kim2021antiferromagnetic}. 

Having established the $J_{\rm{eff}}$ = 1/2 state, we consider the spin model in the hyperoctagon structure. Since all Co ions are crystallographically identical, the nearest neighbor interaction is the same for all Co-Co bonds. In addition, the hyperoctagon lattice is a bipartite lattice. In the simple nearest neighbor Heisenberg model on the 3D bipartite lattice, only a conventional N{\'e}el order is expected at temperature $\sim \Theta$, accompanying a drop of magnetic susceptibility with full entropy release.
In stark contrast, in the phase I', no internal magnetic field is observed and the release of the magnetic entropy is only partial despite the sizable decrease of $\chi$. These features observed in phase I' indicate a development of short-range spin correlations rather than the conventional N{\'e}el order, demonstrating the insufficiency of the Heisenberg model and the presence of a frustration.

Among various origins of the frustration, we first examine the possibility of competing nearest neighbor and further neighbor interactions. In the \ce{Co2(ox)3} network, the second and third neighbor Co-Co distances are 9.5 and 12.2 \AA, respectively. These are even longer than the 5th neighbor Co-Co distance in the typical honeycomb cobaltate \cite{xiao2019crystal}. Interactions between such distant ions are usually neglected. Moreover, the oxalate anions are mutually tilted in the network (Fig.~\ref{fig:fig1}(d)). This tilting will reduce the orbital overlapping for further neighbor interactions. These structural features should strongly suppress the further neighbor interactions in the \ce{Co2(ox)3} network. Therefore, the frustration caused by further neighbor interactions is unlikely.

As a frustration effect realizing the intriguing magnetic state in the phase I’, we propose the emergence of the Kitaev interaction. Kitaev interaction between nearest neighbor ions emerges as a result of the superexchange process via ligand anions \cite{sano2018kitaev, liu2018pseudospin}. In honeycomb cobaltates made of edge-shared \ce{CoO6} octahedra, a dominant Heisenberg interaction from the direct \textit{d}-\textit{d} hopping is suggested due to the short nearest neighbor Co-Co distance $\sim$ 3 \AA \space \cite{winter2022magnetic}. In contrast, the nearest neighbor Co-Co distance in the oxalate network is extended to 5.46 \AA \space by the oxalate anions, suppressing the direct \textit{d}-\textit{d} hopping while maintaining the superexchange paths via oxalate anions. Such a situation should be ideal for the emergence of a dominant Kitaev interaction via the superexchange process, as theoretically proposed in a similar MOF \cite{yamada2017designing}.

In the Kitaev model, the spin degrees of freedom is fractionalized into the two types of quasiparticles related with the itinerant Majorana fermions and $Z_{\rm{2}}$ fluxes \cite{motome2020hunting}. The two quasiparticles have separated energy scales, resulting in the double peaks in the temperature dependence of the specific heat with the release of entropy of the half of \textit{R}ln2 for each peak \cite{nasu2014vaporization,mishchenko2017finite,motome2020hunting,jahromi2021thermodynamics}. An intermediate temperature paramagnetic regime is expected to appear, where the nearest neighbor spin correlations are developed. Our observation of the partial entropy release without long-range order in the phase I' is consistent with the theoretical prediction. On the other hand, the long-range magnetic order below $T_{\rm{L}}$ should be attributed to the presence of non-Kitaev interactions such as a Heisenberg interaction. 
 
The magnetic-field-induced phases can also be explained by the presence of the Kitaev interaction. In the Kitaev-Heisenberg model on the hyperoctagon lattice at zero magnetic field, four magnetically ordered phases, i.e. Néel, zigzag, stripy antiferromagnetic, and ferromagnetic phases, may appear depending on the relative strength between the Kitaev and the Heisenberg interactions \cite{jahromi2021thermodynamics}, similar to the cases in the honeycomb and 3D hyperhoneycomb lattices \cite{chaloupka2013zigzag,fukui2023ground}. Although the magnetic field effect on these phases in the hyperoctagon lattice is yet to be theoretically clarified, the emergence of successive field-induced phases \cite{janssen2016honeycomb} or the metamagnetic behavior \cite{lee2014order} are suggested in the zigzag or stripy antiferromagnetic phases stabilized by the Kitaev interaction in the honeycomb and 3D hyperhoneycomb lattices. This is in sharp contrast to the Néel phase with dominant Heisenberg interaction, where no field-induced anomaly is expected up to the full saturation.

In addition, we suggest different type of frustration, that may be called as bond frustration, is brought by the tilting of the trigonal axis of each \ce{CoO6} octahedra (Fig.~\ref{fig:fig1}(d)). An easy-plane anisotropy perpendicular to trigonal axis can occur in compressed octahedron as observed in various \ce{Co^2+} compounds including honeycomb oxides \cite{xiao2019crystal}, pyrochlore fluorides \cite{ross2017single}, and an organic molecular complex \cite{yao2017anisotropic}. Given the similar compression of the octahedra, a local easy-plane anisotropy is expected in the \ce{Co2(ox)3} network. If the easy-plane anisotropy is strong enough, the line shared by the two mutually tilted easy planes, i.e. Co-Co bond as shown in (Fig.~\ref{fig:fig1}(c)), should become the local easy axis between the adjacent Co ions. Since the trigonal axis directions of all the three surrounding Co ions are different (Fig.~\ref{fig:fig1}(d)), three easy axes appear along the Co-Co bond with the mutual angle of 120$^\circ$. These three easy axes cannot be satisfied simultaneously and give rise to the bond frustration. Note that this frustration is unique to the hyperoctagon structure, in sharp contrast to the honeycomb case in which a global easy-plane anisotropy emerges because all the octahedra are compressed along the same direction.

We suggest that the experimental observations may be compatible with the scenario of the bond frustration. Frustrated magnetism in the cubic 3D lattice with tilted local easy planes, which is similar to the present material, is seen in the XY-pyrochlore magnet \cite{hallas2018experimental}. The double-peak structure in the specific heat is observed experimentally and the field-induced phase transition is proposed theoretically \cite{maryasin2016low,hallas2018experimental}. Our results call for theoretical study on the effect of local spin anisotropy in the hyperoctagon lattice.

\paragraph*{Summary and Perspective}
We present the first experimental realization of the $J_{\rm{eff}}$ = 1/2 hyperoctagon lattice in \ce{[(Me2NH2)3(SO4)]2[Co2(ox)3]}. We find multiple magnetic phases in the \textit{H}\textit{--T} phase diagram, including the intriguing phase I' without long-range magnetic order, indicating the presence of a frustration in the bipartite lattice. As a possible origin of the frustration, we propose the Kitaev interaction in the nearest neighbor superexchange process and/or a bond frustration caused by the local magnetic anisotropy. Further investigations including the first-principles calculations and neutron scattering experiments would significantly advance the understanding of the frustrated magnetism in the spin-orbital coupled system. Investigations on the relative MOFs are also of interest \cite{hernandez1998weak,nakashima2023enhanced}. Structural parameters can be modified by introducing other counter molecules, resulting in the changes in the magnetic interactions and the ground state. Moreover, long Co-Co distances in the oxalate MOF may allow the search for the pressure-induced magnetic phases, which is often hindered by the dimerization in 4\textit{d} and 5\textit{d} Kitaev materials \cite{bastien2018pressure}.

\begin{acknowledgments}
This work was supported by JSPS KAKENHI Grants No. JP22H04467, JP22K13996, and JP23H01116.
\end{acknowledgments}

\bibliography{manuscript}

\end{document}